\documentclass[journal]{IEEEtran}
\usepackage{amsmath, amssymb, amsfonts}
\usepackage{algorithmic}
\usepackage{algorithm}
\usepackage{array}
\usepackage[caption=false,font=scriptsize,labelfont=sf,textfont=sf]{subfig}
\usepackage{textcomp}
\usepackage{stfloats}
\usepackage{url}
\usepackage{verbatim}
\usepackage{cite}
\usepackage{bm}

\usepackage{tabularx, multirow,multicol,colortbl,array,adjustbox}
\usepackage{graphicx,color, array}
\usepackage[dvipsnames]{xcolor}
\usepackage{float}
\usepackage{comment,psfrag}
\usepackage{enumitem}
\usepackage{tikz,tikzscale,bigints,pgfplots}
\pgfplotsset{compat=1.18}
\usetikzlibrary{positioning, arrows.meta, colorbrewer, calc, fit, matrix,external,patterns}
\usepgfplotslibrary{fillbetween}
\usepackage{multirow}

\begin{document}

\title{Physics-Aware Initialization Refinement in Code-Aided EM for Blind Channel Estimation}

\author{Chin-Hung Chen,~\IEEEmembership{Student Member,~IEEE,}
Ivana Nikoloska,~\IEEEmembership{Member,~IEEE,}
\\Wim van Houtum,~\IEEEmembership{Senior Member,~IEEE,}
Yan Wu, and Alex Alvarado,~\IEEEmembership{Senior Member,~IEEE}
\thanks{Manuscript received xx, xxxx; revised xx, xxxx.}
\thanks{C.-H. Chen, I. Nikoloska, W. van Houtum, and A. Alvarado are with the Information and Communication Theory (ICT) Lab, Electrical Engineering Department, Eindhoven University of Technology, 5600MB Eindhoven, The Netherlands. \textit{(e-mails: c.h.chen@tue.nl; i.nikoloska@tue.nl; a.alvarado@tue.nl)}.} 
\thanks{W. van Houtum and Y. Wu are with NXP Semiconductors, High Tech Campus 60, 5656AE Eindhoven, The Netherlands. \textit{(e-mails: wim.van.houtum@nxp.com; yan.wu\_2@nxp.com)}.}

}

\maketitle

\begin{abstract}
    This paper addresses the well-known local maximum problem of the expectation-maximization (EM) algorithm in blind inter-symbol interference (ISI) channel estimation. This problem primarily results from phase and shift ambiguity due to poor initialization, which a blind EM estimation is inherently unable to distinguish. We propose an effective initialization refinement algorithm that utilizes the decoder output as a metric for model selection. Finite candidate models are generated based on the physical properties of the channel and modulation format, incorporating a joint detection of phase and shift ambiguities. Our results show that the proposed algorithm significantly reduces the number of local maximum cases to nearly one-third for a 3-tap ISI channel under highly uncertain initial conditions. The improvement becomes more pronounced as initial errors increase and the channel memory grows. When used in a turbo equalizer, the proposed algorithm is required only in the first turbo iteration, which limits any complexity increase with subsequent iterations.

\end{abstract}

\vspace{-3mm}
\section{Introduction}\label{sec:intro}
    Inter-symbol interference (ISI) is a central problem for wireless communications. To accurately estimate the channel response without transmitting extra training symbols, the expectation-maximization (EM) algorithm has become a popular technique for blind channel estimation \cite{Ghosh92, Kaleh94, Anton97}. To further improve EM estimators, code-aided systems have been introduced. By iteratively refining the channel estimate using extrinsic information from error correction decoders, such as convolutional codes \cite{Lopes01, Garcia03, Zhao10, CHC25}, or low-density parity check (LDPC) codes \cite{Niu05, Gunther05}, the estimation improves significantly compared to standalone EM. However, a fundamental limitation of the EM algorithm is its susceptibility to poor initializations, which can misguide the E-step of the EM algorithm, particularly in the early iterations. This problem will eventually cause convergence to a local maximum and severely degrade estimation performance \cite{Lopes01_2, Garcia03, Zhao10, CHC25, Niu05, Gunther05, Yang12}.

Key factors that cause local convergence in EM-based estimation include phase and shift ambiguities that result from poor initialization.
Phase ambiguity is commonly encountered in blind estimation of phase-shift keying (PSK) signals due to the inherent symmetries in the constellation. This issue has been discussed in \cite{Garcia03, Lopes01_2}, but a rigorous solution has yet to be proposed. While strategies, such as differential PSK, have been used to circumvent this issue \cite{XMChen01, Serbetli14}, an effective phase detection algorithm for PSK remains an open challenge. Meanwhile, shift ambiguity arises from the convolutional invariance property, where a circular shift in the channel tap order leads to the same set of Gaussian means that the EM algorithm cannot resolve \cite{Anton97, XMChen01}. 

To achieve reliable initialization, advanced methods such as semi-blind (data-aided) EM \cite{Kutz10, Nayebi18, Yang20} and deep learning–aided EM \cite{Schmid25} have been proposed. These approaches rely on additional pilot symbols for data-driven techniques to adjust the initialization value within an acceptable error range, thereby preventing the EM algorithm from diverging. However, in fully blind systems, such pilots are inherently unavailable, limiting their applicability.

This paper proposes a novel joint blind phase and shift ambiguity detection algorithm for fully blind receivers. Unlike existing approaches, the proposed method operates solely on the received data without requiring pilot symbols. The key idea is to harness the inherent symmetric properties of communication systems. Specifically, the convolutional invariance of the linear ISI channel and the rotational invariance of PSK modulation enable the construction of a finite set of physics-consistent models. By restructuring the extrinsic likelihood function from the EM-based estimator-equalizer, the decoder resolves phase and shift ambiguities by selecting the candidate that maximizes the corresponding model evidence. The algorithm is applied only in the first turbo iteration to correct ambiguities arising from poor initialization. Simulation results demonstrate that this physics-aware initialization refinement significantly improves convergence stability and detection performance across various channel models and lengths.

\section{System Model and Receiver Design}\label{sec:sys}
    \begin{figure*}
    \vspace{-10mm}
    \centering
    \resizebox{.93\textwidth}{!}{\includegraphics{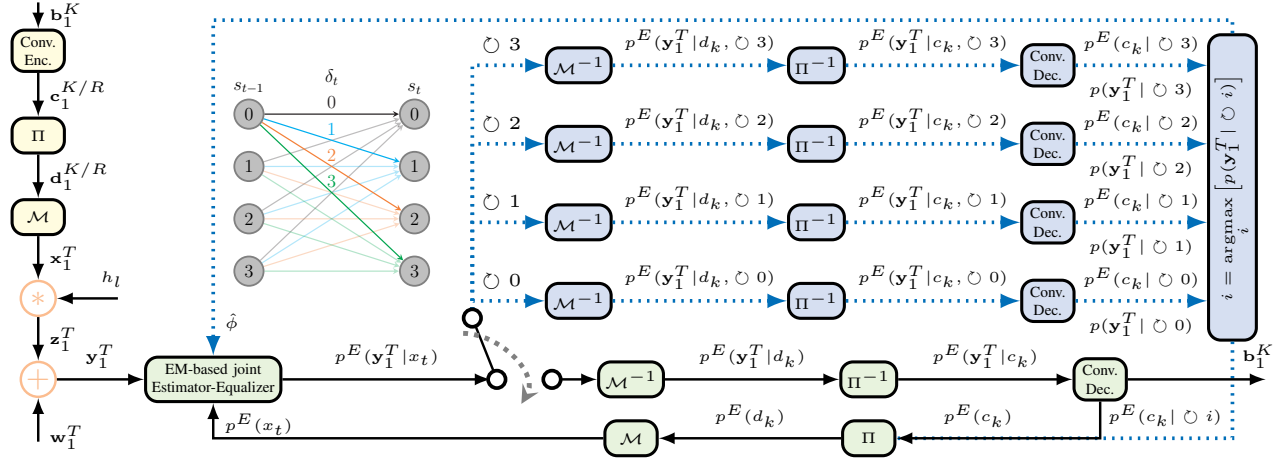}}
    \caption{Schematic diagram of the considered communication system. The bit-interleaved coded modulation for transmission over a linear ISI and AWGN channel is represented by the yellow blocks. The conventional joint estimation and turbo equalization receiver is depicted in green blocks. The blue blocks represent the initialization refinement technique for QPSK modulation based on the proposed phase ambiguity detection algorithm.}
    \label{fig:sys_block}
    \vspace{-5mm}
\end{figure*}

This section details the system model, EM-based estimator-equalizer, joint estimation with turbo equalization, and the phase and shift ambiguity detection algorithm with the overall system design shown in Fig.~\ref{fig:sys_block}. 


\subsection{Transmission Over Linear ISI And AWGN Channel}
A convolutional encoder takes as input the information bit sequence $\mathbf{b}_1^K = (b_1, b_2, \cdots, b_{K})$ and generates a coded bit sequence $\mathbf{c}_1^{K/R}$, where $R$ represents the coding rate. An interleaver ($\Pi$) is used to produce the interleaved coded bit sequence $\mathbf{d}_1^{K/R}$. Then, an $M$-PSK symbol mapper ($\mathcal{M}$) maps the binary coded bits to symbols $x_t \in \mathcal{X}=\{e^{j2\pi i /M} \mid i=0,1,\dots,M-1 \}$, where $M=|\mathcal{X}|$. The symbol time step is denoted by $t$, and the length of the symbol sequence is $T = K/R/\log_2M$. The block-wise invariant linear ISI and AWGN channel can be modeled as
\begin{align}\label{eq:channel_mod}
    y_t = \sum_{l=0}^{L-1} h_{l} x_{t-l}  + w_t = z_t + w_t, \quad w_t \sim \mathcal{CN}(0,\sigma_w^2).
\end{align}
The noise realization $w_t$ is drawn from a complex Gaussian with zero mean and variance $\sigma_w^2$. We use $z_t \in \mathcal{Z}= \{\mu_0,\mu_1,\ldots,\mu_{E-1}\}$ to represent the noiseless ISI channel output. The elements in $\mathcal{Z}$ are the $E=M^L$ possible outcomes of convolving the PSK symbol with the channel taps.

\subsection{EM-Based Estimator-Equalizer}
The optimal MAP equalizer computes the posterior probability of the transmitted symbol $x_t$ as
\begin{align}\label{eq:jointx}
    p(x_t|\mathbf{y}_1^T) \propto p(x_t,\mathbf{y}_1^T) = \sum_{x_t:(s_{t-1} \rightarrow s_{t})}{ {p(s_t, s_{t-1}, \mathbf{y}_1^T)}},
\end{align}
where $s_t$ denotes the state realizations drawn from a time-invariant finite-state space $\mathcal{S}=\{0,1,\cdots,S-1\}$ and $x_t:(s_{t-1} \rightarrow s_{t})$ denotes all transitions $(s_{t-1} \rightarrow s_{t})$ that are driven by the symbol $x_t$. The joint probability $p(s_t, s_{t-1}, \mathbf{y}_1^T)$ on the right-hand side of \eqref{eq:jointx} can be computed via
\begin{align} \label{eq:map2}
    p(s_t, s_{t-1}, \mathbf{y}_1^T)=\underbrace{p(\mathbf{y}_{t+1}^{T} | s_{t})}_{\beta(s_t)} \, \underbrace{p(y_t, s_t | s_{t-1})}_{\gamma(s_t,s_{t-1})} \, \underbrace{p(s_{t-1}, \mathbf{y}_1^{t-1})}_{\alpha(s_{t-1})},
\end{align}
\begin{align} 
    \alpha(s_{t}) &= p(s_{t}, \mathbf{y}_1^{t}) = \sum_{s_{t-1}\in\mathcal{S}} \gamma(s_{t}, s_{t-1}) \cdot \alpha(s_{t-1}), \label{eq:siso_alpha}  \\ 
    \beta(s_{t-1}) &= p(\mathbf{y}_{t}^{T} | s_{t-1}) = \sum_{s_{t}\in\mathcal{S}} \beta(s_t) \cdot \gamma(s_{t}, s_{t-1}),  \label{eq:siso_beta} \\ 
    \gamma(s_{t}, s_{t-1}) &= p(y_t, s_t | s_{t-1}) = p(y_t| s_t, s_{t-1})p(s_t|s_{t-1}). \label{eq:gamma}
\end{align}
with $\alpha$, $\beta$, and $\gamma$ denote the forward, backward recursions, and branch metric. The computations from \eqref{eq:jointx} to \eqref{eq:siso_beta} form the well-known BCJR algorithm.

Following \cite{Kaleh94,CHC25}, we define the total number of states as $S=M^{L-1}$. This forms a set of edges $\delta_t \in \{0,1,\ldots,E-1\}$ with each edge associated with two consecutive states $\delta_t=l  \iff(s_{t-1}=i \rightarrow s_t=j)$ and is used as an indicator to identify which Gaussian component $\mathcal{CN}(\mu_l, \sigma_w^2)$ produces the corresponding observed data $y_t$. A trellis example for a quadrature PSK (QPSK) with an $L=2$ channel is shown in the inset in Fig.~\ref{fig:sys_block}. The likelihood function and the transition probability on the right-hand side of \eqref{eq:gamma} are
\begin{align}\label{eq:lik_modem}
    p(y_t|s_t,s_{t-1}) = p(y_t | \delta_t=l) = \frac{1}{{\pi \sigma_w^2}}\exp{\left(-\frac{|y_t-{\mu}_l|^2}{{\sigma}^2_w}\right)},  
\end{align}
\begin{align} \label{eq:state_tran}
    p(s_t|s_{t-1}) = p(\delta_t=l)= 
    \begin{cases}
        p(x_t), & x_t: (\delta_t=l) \\
        0, &\text{otherwise}.
    \end{cases}
\end{align}
To obtain accurate channel likelihood \eqref{eq:lik_modem}, channel parameters $\mu_l$ and $\sigma^2_w$ are required. This paper focuses primarily on estimating $\mu_l$ with the variance $\sigma^2_w$ assumed to be known. Therefore, the parameter set that we aim to estimate is $\Theta^{(n)} = \{\hat{\mu}_{l}^{(n)}\}$ with superscripts ${(n)}$ denoting the $n$-th iteration. The EM algorithm performs blind estimation through two iterative steps: the E-step and the M-step. With $\Theta$ held fixed, the E-step computes the posterior belief of each edge via 
\begin{align} \label{eq:pos_st}
    p(\delta_t=l | \mathbf{y}^T_1, \Theta^{(n)}) = \frac{p(\delta_t=l, \mathbf{y}^T_1| \Theta^{(n)})}{\sum_{\delta_t }p(\delta_t,  \mathbf{y}^T_1| \Theta^{(n)})},
\end{align}
where $p(\delta_t,\mathbf{y}_1^T|\Theta^{(n)})=p(s_{t}, s_{t-1}, \mathbf{y}^T_1|\Theta^{(n)})$ represents the joint probability of the edges $\delta_t$ and the complete observation $\mathbf{y}_1^T$, which can be efficiently calculated via the BCJR algorithm in \eqref{eq:map2}--\eqref{eq:state_tran}.
The M-step conducts the maximum likelihood estimation \cite{Bishop} of the model parameter $\Theta$ based on \eqref{eq:pos_st} via
\begin{align}
    \tilde{\mu}_l^{(n+1)} = \frac{\sum_{t=1}^{T} p(\delta_t=l | \mathbf{y}_1^T, \Theta^{(n)}) \cdot y_t} {\sum_{t=1}^{T} p(\delta_t =l | \mathbf{y}_1^T, \Theta^{(n)})}. \nonumber
\end{align}
To further improve the estimation for the linear ISI model \eqref{eq:channel_mod}, we apply a linear constraint \cite{Anton97} to the Gaussian means as
\begin{align}
    &\left[\hat{\mu}_0^{(n+1)},\hat{\mu}_1^{(n+1)},\ldots,\hat{\mu}_{E-1}^{(n+1)}\right]^T = \mathbf{D} \hat{\mathbf{h}}, \nonumber \\
    &\hat{\mathbf{h}}=(\mathbf{D}^T\mathbf{D})^{-1}\mathbf{D}^T\left[\tilde{\mu}_0^{(n+1)},\tilde{\mu}_1^{(n+1)},\ldots,\tilde{\mu}_{E-1}^{(n+1)}\right]^T, \label{eq:linear}
\end{align}
here $\hat{\mathbf{h}}=[\hat{h}_0,\hat{h}_1,\ldots,\hat{h}_{L-1}]^T$ is the vector of channel taps and $\mathbf{D}$ is an $E\times L$ matrix with its row vector corresponding to the $L$ consecutive PSK symbols associated with the $E$ different ISI noiseless channel outputs.

The EM algorithm then advances to the next iteration by using the updated model parameters set $\Theta^{(n+1)} = \{\hat{\mu}_{l}^{(n+1)}\} $. Once the specified EM iterations $N$ is satisfied, the joint symbol probability based on the final estimated parameter $\Theta^{(N)}$ is computed via \eqref{eq:jointx} with the likelihood function \eqref{eq:lik_modem} calculated based on the final mean estimation $\hat{{\mu}}_l^{(N)}$.
\subsection{Code-Aided Joint Estimation And Turbo Equalization}\label{sec:codejoint}
The light green blocks in Fig.~\ref{fig:sys_block} show the conventional code-aided joint estimation and turbo equalization design, where the extrinsic information $p^E(x_t)$ provided by the decoder is used to update the transition probability \eqref{eq:state_tran}. The extrinsic information refers to the information provided by one decoder (equalizer) about a specific bit (symbol), excluding the information used to derive it directly. We first compute the extrinsic information from the equalizer as
\begin{align} \label{eq:extx}
     p^E(\mathbf{y}_1^T|x_t) = {p(x_t, \mathbf{y}_1^T)}/{p^E(x_t)},
\end{align}
where we set $p^E(x_t) = 1 / M$ in the first turbo iteration.
We first perform symbol-to-bit demapping $p^E(\mathbf{y}_1^T | d_k) = \mathcal{M}^{-1}[p^E(\mathbf{y}_1^T| x_t)]$, followed by de-interleaving $p^E(\mathbf{y}_1^T|c_{k}) = \Pi^{-1} \left[ p^E( \mathbf{y}_1^T | d_k) \right]$. This extrinsic information of coded bits is then fed to the convolutional decoder to generate the joint probability $p(\tilde{s}_k, \tilde{s}_{k-1}, \mathbf{y}_1^T)$ using the BCJR algorithm \eqref{eq:map2}--\eqref{eq:siso_beta} with $\tilde{s}_k \in \{0,1,\ldots,2^{L_c}-1\}$ defined as the decoder state with $L_c$ denoting the constraint length of the convolutional code. The joint probability of the coded bits $c_k$ is then computed as
\begin{align}
    p(c_k, \mathbf{y}_1^T) = \sum_{c_k:(\tilde{s}_{k-1} \rightarrow \tilde{s}_{k})}{ {p(\tilde{s}_k, \tilde{s}_{k-1}, \mathbf{y}_1^T)} }, \label{eq:jointc}
\end{align}
followed by the extrinsic information computation via
\begin{align} 
    p^E(c_k) = {p(c_k,\mathbf{y}_1^T)}/{p^E(\mathbf{y}_1^T|c_k)}.  \nonumber
\end{align}
After bit interleaving 
$p^E(d_k) = \Pi \left[ p^E(c_k) \right]$, the extrinsic prior symbol information is derived via a bit-to-symbol mapping
\begin{align}\label{eq:ext_x}
    p^E(x_t) = \mathcal{M}[p^E(d_k)].
\end{align}
Finally, \eqref{eq:ext_x} is fed back to the estimator-equalizer to replace $p(x_t)$ on the right-hand side of \eqref{eq:gamma} in the next turbo iteration. Note that we refer to the turbo iteration as the entire joint estimation and turbo equalization loop. In contrast, the EM iteration refers to the optimization within the EM estimator.

\subsection{Physics-Aware Initialization Refinement Algorithm}
In a symmetric constellation like $M$-PSK, $M$ possible phase shifts result in the same constellation shape, leading to phase ambiguity. Shift ambiguity, on the other hand, arises from convolutional invariance. This means that $h_l$ and $h_{l-\tau}$ produce the same set of Gaussian means $\mathcal{Z}$. Consequently, EM-based blind estimators cannot effectively address these ambiguities. The core idea of our proposed initialization refinement algorithm is to use the decoder output for model selection, allowing us to identify and resolve these physical ambiguities.

We first present the phase ambiguity detection algorithm shown in the blue blocks in Fig.~\ref{fig:sys_block}. The intuition behind our proposal is to use the decoder's sensitivity to the symbol-to-bit mapping to detect phase ambiguity effectively. We first permute the order of \eqref{eq:extx} and generate $M$ possible extrinsic symbol information as 
\begin{align}\label{eq:phs_ext}
    p^E(x_t\circlearrowright i, \mathbf{y}_1^T), \quad i=\{0,1,\ldots,M-1\},
\end{align}
where we use $\circlearrowright$ to denote the circular shift operation.
This information is then passed onto $M$ parallel demappers ($\mathcal{M}^{-1}$), deinterleavers ($\Pi^{-1}$), and decoders as described in Sec.~\ref{sec:codejoint} to derive $p(c_k,\mathbf{y}_1^T|\circlearrowright i)$ in \eqref{eq:jointc}. The model evidence is then computed for each of the parallel processing modules as
\begin{align}\label{eq:phs_evi}
    p(\mathbf{y}_1^T|\circlearrowright i) = \sum_{c_k}p(c_k, \mathbf{y}_1^T |\circlearrowright i),
\end{align}
where we determine the phase ambiguity via
\begin{align}
    \hat{\phi} = {2\pi i/M}, \quad i=\operatorname*{argmax}\limits_{i\in \{0,1,\dots,M-1\}}{\left[p(\mathbf{y}_1^T|\circlearrowright i)\right]}. \nonumber
\end{align}

To attack the shift ambiguity problem, we make use of the intermediate channel tap estimates in \eqref{eq:linear} and circular shift the channel taps estimation $\hat{\mathbf{h}}\circlearrowright \tau$ with $\tau=0,1,\ldots,L-1$. This operation leads to $L$ distinct extrinsic symbol information $p^E(x_t, \mathbf{y}_1^T|\hat{\mathbf{h}}\circlearrowright \tau)$ calculated via \eqref{eq:jointx}. These metrics are forwarded to the phase detection block, along with \eqref{eq:phs_ext}, resulting in $LM$ possible models that are then passed onto \eqref{eq:phs_evi} to jointly determine the phase $\hat{\phi}$ and shift $\hat{\tau}$. These values will be used to refine the channel estimates once the EM estimator-equalizer completes its $N$th EM iterations as
\begin{align}
    &\left[\hat{\mu}_0^{(N)},\hat{\mu}_1^{(N)},\ldots,\hat{\mu}_{E-1}^{(N)}\right]^T = e^{-j\hat{\phi}}\cdot\mathbf{D} (\hat{\mathbf{h}}^{(N)}\circlearrowright \hat{\tau}). \nonumber 
\end{align}

\subsection{Computational Complexity}
We compare the complexity (big-$\mathcal{O}$) of the conventional code-aided and the proposed physics-aware EM systems, focusing on the dominant multiplications in the BCJR operation (E-step and the MAP decoder). For a channel memory $L$, the ISI trellis has $M^{L-1}$ states and $M$ outgoing branches per state, i.e., $M^{L}$ branches per trellis section. Over $T$ sections, the E-step thus incurs $\mathcal{O}\!\left(TM^{L}\right)$ multiplications. In each turbo iteration, the EM is first executed $N$ times, yielding $\mathcal{O}(N T M^{L})$. The decoder processes $T \log_2 M$ coded bits with $2^{L_c-1}$ states, resulting in $\mathcal{O}\!\left(T\log_2 M \cdot 2^{L_c}\right)$ multiplications. Hence, the total cost per turbo iteration is $\mathcal{O}\!\left(N T M^{L} + T\log_2 M \cdot 2^{L_c}\right)$. To resolve phase and shift ambiguities, $M$ phase rotations and $L$ circular shifts are evaluated once in the first turbo iteration. Each hypothesis reuses the decoder, giving an additional one-time cost $\mathcal{O}\!\left((L-1)(M-1) \,\big[T\log_2 M \cdot 2^{L_c}\big]\right)$.

\section{Simulation Results}\label{sec:simu}
\begin{figure*}[t]
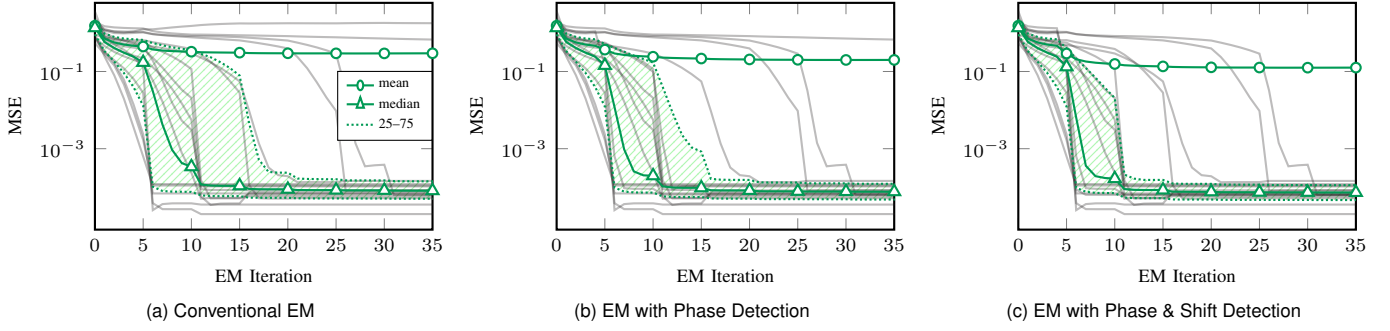

    \vspace{-5pt}
    \subfloat[Conventional EM\label{fig:e05_std}]{%
        \centering
        \includegraphics[width=.33\textwidth]{fig/Err_e05_std.tikz} 
    }
    \subfloat[EM with Phase Detection\label{fig:e05_phs}]{%
        \centering
        \includegraphics[width=.33\textwidth]{fig/Err_e05_phs.tikz} 
    }
    \subfloat[EM with Phase \& Shift Detection\label{fig:e05_joi}]{%
        \centering
        \includegraphics[width=.33\textwidth]{fig/Err_e05_joi.tikz}  
    }
    \caption{Mean square error of the estimated mean. We consider the $L=3$ channel with an initialization error of $\sigma^2_h=0.5$ for conventional EM, EM with phase detection, and EM with joint phase and shift detection algorithms. Gray lines indicate individual realizations (15 out of 2000). The mean and median of the 2000 independent simulations are represented with circle and triangle solid green lines, respectively, while the shaded areas indicate 25--75 percentiles.}
    \vspace{-2mm}
\end{figure*}

\begin{figure}[t]
    \centering
    \includegraphics[width=.88\columnwidth,height=.65\columnwidth]{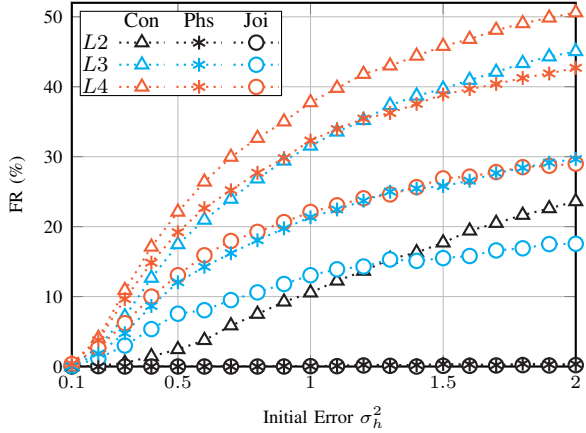} 
    \caption{Failure rate of the EM estimators (conventional, phase detection, and joint detection) after $7$ turbo iterations ($35$ EM iterations) concerning initial condition errors $\sigma_h^2$ for channel memory $L=2$, $3$, and $4$.}
    \label{fig:fail}
    \vspace{-3mm}
\end{figure}

All simulation results are obtained from $2000$ independent Monte Carlo runs. A rate-$1/2$ convolutional code with generator polynomials $(5,7)_8$ is applied with $2$ zero-padding bits to terminate the trellis. A random bit-interleaver of the same length as the codeword is used, and the interleaved bits are mapped to QPSK symbols. 
While the proposed algorithm consistently achieves better performance than the baseline, the improvement becomes more evident when the block length exceeds $1000$ information bits for the specific code used in this study. Accordingly, each frame in subsequent simulations is set to $10000$ information bits to highlight its effectiveness.
We evaluate our proposed algorithm with three different length ($L=2,3,4$) block-wise invariant ISI channels with coefficients $(e^{j\theta_1},-e^{j\theta_2})$,  $(0.5e^{j\theta_1}, 0.7e^{j\theta_2},0.5e^{j\theta_3})$, and $(0.38e^{j\theta_1}, 0.6e^{j\theta_2},0.6e^{j\theta_3}, 0.38e^{j\theta_4})$, respectively. The channel phase is drawn from a uniform distribution $\theta_l\sim\mathcal{U}(0,2\pi)$. The channel gains for $L=3,4$ are selected based on their worst-case minimum Euclidean distance at the output \cite{Proakis}, which enables us to evaluate our proposed algorithm in the worst ISI scenarios. We set $\text{SNR}=||\mathbf{h}||^2 E\{|x_t|^2\}/\sigma_w^2$ equal to $6$~dB, where we assume the noise variance $\sigma^2_w$ is known. The system is initialized with $\hat{h}_l^{(0)}=h_l + \epsilon_l$ where $\epsilon_l\sim \mathcal{CN}(0,\sigma^2_h)$. In each turbo iteration, the estimator performs $5$ internal EM iterations before passing the estimated likelihoods to the turbo equalization module. Note that phase and shift ambiguity detection is only performed at the first turbo iteration (i.e., after the first $5$~EM iteration). We use two metrics for evaluating our system performance: the mean square error (MSE) of the estimated mean $\text{MSE}=\frac{1}{L}\sum_{l=1}^{L}|\hat{\mu}_l-\mu_l|^2$ and the failure rate (FR), defined as the percentage of runs where the square error exceeds $10^{-1}$ after $7$ turbo iterations (i.e., $35$ EM iterations). 
\footnote{The optimal estimation achieves $\text{MSE}=10^{-4}$ at $\text{SNR}=6$ dB \cite{CHC25}; thus, $\text{MSE}=10^{-1}$ offers a clear margin for identifying local maxima.} 
To ensure the reliability of the detected phase and shift ambiguity, we only perform the initialization refinement if the logarithmic evidence of the selected model $\ln{p(\mathbf{y}_1^T|\circlearrowright i)}$ is $10^3$ larger than the rest of the models.

In Figs.~\ref{fig:e05_std}--\ref{fig:e05_joi}, we show the MSE of the mean estimation for an $L=3$ channel. A significant difference exists between the mean and median, mainly due to the outliers represented by the gray lines above $10^{-1}$, a phenomenon also identified in \cite{Schmid25}. Using the phase detection technique, the variance is markedly reduced compared to the conventional EM, as clearly shown by the reduced size of the green-shaded area in Fig.~\ref{fig:e05_phs}. This improvement is further enhanced by incorporating the joint phase and shift detection, where most of the outliers are corrected at the $5${th} EM iteration (Fig.~\ref{fig:e05_joi}). In Fig.~\ref{fig:fail}, the FR is reported concerning the initialization error $\sigma_h^2$. The results indicate that our proposed phase detection technique can reduce the FR from $45$\% to $30$\% for the $L=3$ channel with an error variance $\sigma^2_h=2$. The FR is further reduced to nearly one-third of the conventional EM by using the joint detection algorithm. The longer the memory, the greater the improvement achieved through the joint detection algorithm, as the chance of shift ambiguity increases, making phase detection alone insufficient. For channels with memory $L=2$, it is sufficient to use only the phase detection algorithm, which adequately addresses any potential shift ambiguity.

\section{Conclusions}\label{sec:conc}
This paper proposes an effective initialization refinement algorithm that corrects phase and shift ambiguities, which commonly cause local maximum problems in EM blind channel estimation. By modifying the soft-output decoder to output model evidence, the detection algorithms select the metric that maximizes the model evidence to detect the phase and shift ambiguities. The proposed algorithm has been evaluated under various levels of initial condition uncertainty and in three different length ISI channel models. Our simulation demonstrates that the proposed joint phase and shift detection algorithm can reduce the failure rate to one-third that of the conventional EM estimator. As the channel memory and modulation order increase, the complexity grows exponentially and may limit the performance of our proposed algorithm. Extending the framework to OFDM could alleviate this limitation, while further generalization to time-varying channels and other modulation formats, such as QAM, would enhance its robustness and applicability to modern wireless systems.


\vfill

\end{document}